\documentclass[a4paper,11pt]{article}
\usepackage{pos}
\usepackage{caption}
\usepackage{subcaption}

\title{Future measurements of TCS at JLab Hall C}

\author*[a]{Debaditya Biswas}
\author[a]{Marie Bo\"er}

\affiliation[a]{Virginia Tech,\\
  Blacksburg, USA}

\emailAdd{debadityab22@vt.edu}
\emailAdd{mboer@vt.edu}

\abstract{Generalized parton Distributions (GPDs) are important functions to understand the three dimensional structure of the nucleon. Deeply Virtual Compton Scattering is one of the 
reaction accessing GPDs, and has been measured for the past $\sim$20 years. However, to move forward, we need to look for other reactions, such as   Timelike Compton Scattering (TCS), its "time-reversal" equivalent. Indeed, accessing GPDs from both DVCS and TCS independently will allow us, for instance, to study their universality. Any assesment on GPD's universality would be a milestone in our field.  
In this article we discuss our preliminary studies on the feasibility of measuring unpolarized and beam polarized cross sections and beam spin asymmetry for TCS
 in the dilepton photoproduction reaction. For that purpose, we use a polarized photon beam and an unpolarized target at JLab Hall C. We will discuss our Geant4 simulations, with a dedicated detector setup along with the use of the SBS magnet for separating outgoing $e^{+}$, $e^{-}$ pairs.}

\FullConference{%
  25th International Symposium on Spin Physics,\\
  24-29 September 2023\\
  Durham, NC, USA
}

\begin{document}
\maketitle

\section{Introduction}
\begin{figure}[!htb]
\centering
\includegraphics[width=0.7\textwidth,angle=0]{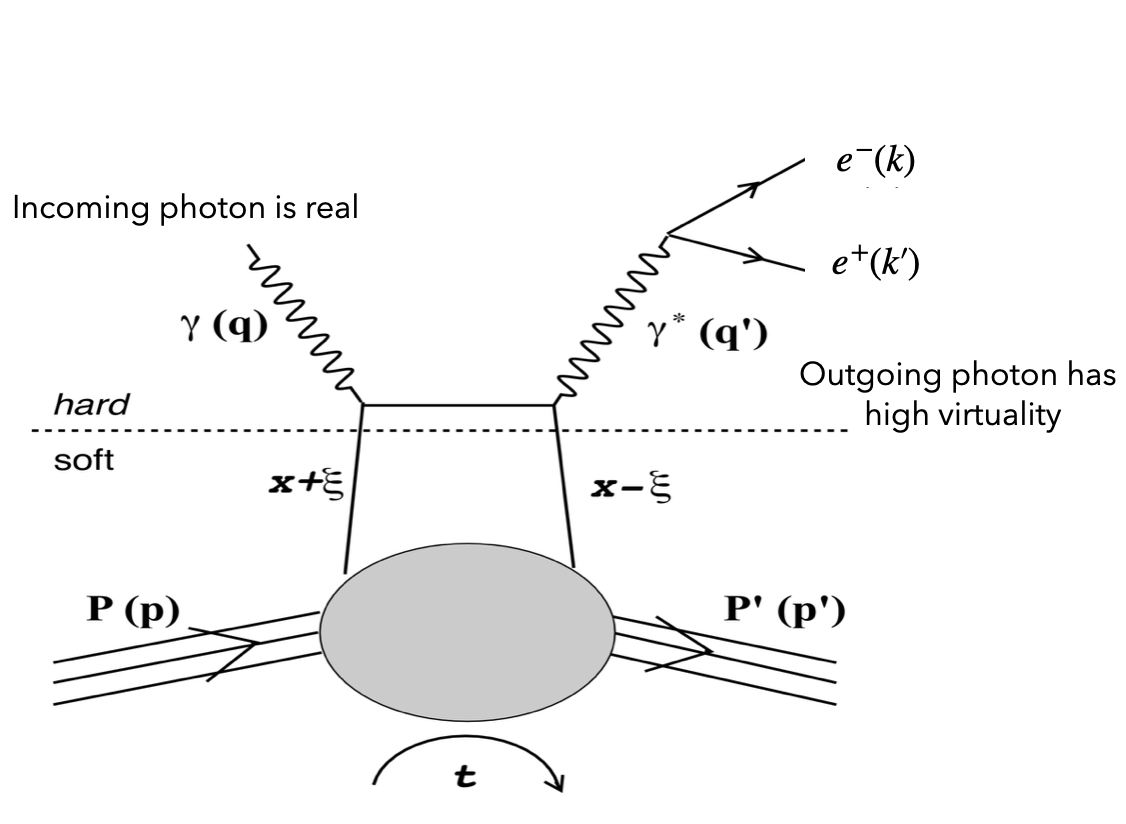}
\caption{Time Like Compton Scattering (leading order and leading twist, CC not represented).}
\label{fig:unpol_tcs_reaction}
\end{figure}

This article discuss our  preliminary studies for measuring  Timelike Compton Scattering (TCS, displayed  Fig~\ref{fig:unpol_tcs_reaction}) at Jefferson Lab Hall C, off an unpolarized Liquid Hydrogen target and a circularly polarized high intensity photon beam. Measuring TCS cross sections and beam spin asymmetries with an unpolarized target, is very important to constrain the   Generalized Parton Distribution (GPDs)~\cite{Mul94,Ji:1996nm} . GPDs are sensitive to the longitudinal momentum versus transverse position structure of partons (quarks and gluons) in  the nucleon~\cite{Die03}. They can't be accessed directly from experiments: we are actually measuring functions of GPDs, called 
 Compton Form Factors (CFFs). There are several GPDs corresponding to different possible relative orientations of the helicity of particles involved in the reaction~\cite{Guidal:2013rya}. 
 In our case, 
 the observables we aim to measure are most sensitive to the GPD "H", the one which is insensitive to the quark and to the nucleon's helicities. GPD H is 
 currently well constrained from DVCS measurements, and this is why we would like to obtain similar measurements from TCS, for a comparison, and for universality studies. 
Furthermore, GPDs are real functions, but CFFs that we are measuring are complex functions: from DVCS measurements, we better constrain the imaginary part of the CFFs. Our equivalent measurement with TCS is sensitive to both real and imaginary parts, and thus will bring more constraints on the real part of the amplitudes in a multi-channel CFF extraction approach. We refer to articles~\cite{Ber02,Boe15} for the phenomenology of TCS off the proton and projections of observables. 

TCS is measured in the reaction $\gamma P \to e^+ e^- P'$, where P is a nucleon (a proton here) and e is a lepton (an electron here). It interferes with another process, called Bethe-Heitler
 (BH) where the lepton pair is produced by a splitting of the incoming photon in the nucleon's field. BH is insensitive to the GPDs. It is parametrized by Form Factors. 
 What we want to measure ("observables") is cross sections ($\sigma$) and  beam spin asymmetries ($A_{\odot U}$ is mostly sensitive to the interference term in the TCS+BH amplitude). The beam spin asymmetry is defined as: 
\begin{equation}
    A_{\odot U} = \frac{\sigma^{+}-\sigma^{-}}{\sigma^{+}+\sigma^{-}},
\end{equation}
where $\odot$ represents the beam polarization and $U$ represents the unpolarized target. $+$, $-$ represents the right and left circularly polarized beam respectively. $\sigma$ is defined as the five differential  polarized cross section, which reads
\begin{equation}
    \sigma^{\pm} \equiv \frac{d^{5}\sigma}{dQ'^2 \ dt \ d (cos \theta) \ d\phi \ dE_{\gamma}},
\end{equation}
 where $Q'^2$ is the virtuality of the outgoing photon, $t$ is the squared momentum transfer (Mandelstam variable), 
 $\theta$ and $\phi$ are the polar and azimuthal angles for the scattered electron in the virtual photon's rest frame, relative to the proton-incoming photon's frame.
 $E_{\gamma}$ is the incoming beam energy.  \\

 In this paper, we will focus on our experimental setup for the measurement of unpolarized and beam polarized TCS+BH
 cross sections at JLab Hall C. Our setup is strongly inspired by a similar one, proposed to measure   TCS+BH transverse target spin asymmetries off an ammonia target~\cite{PolarizedTCS}.

\section{Experimental Setup}

\begin{figure}[!htb]
\centering
\includegraphics[width=1.0\textwidth,angle=0]{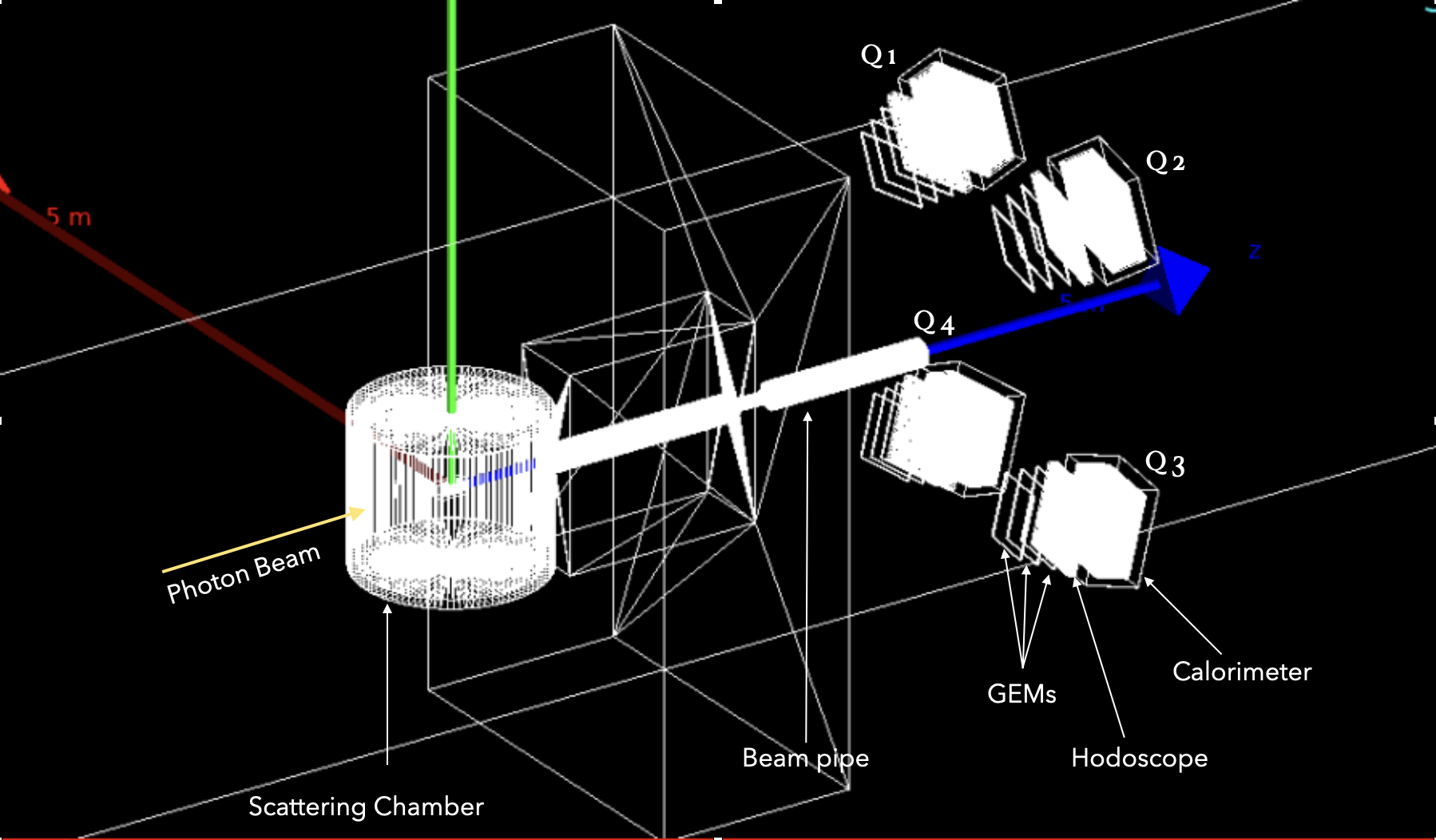}
\caption{The GEANT4 simulations for the  experimental setup for an unpolarized TCS experiment. Q1, Q2, Q3 and Q4 designate four  quadrants consisting of the same set of detectors. Each of the four quadrants consists of three layers of GEMs, 3 layers of hodoscopes and an electromagnetic calorimeter. The unpolarized liquid hydrogen target is inside the scattering chamber (on the left). The SBS magnet is shown in between the "quadrants" and the scattering chamber.}
\label{fig:unpol_tcs_setup_full}
\end{figure}

To measure unpolarized TCS cross sections at JLab Hall C, we need a dedicated experimental setup. 
A lot of studies were done for the equivalent polarized TCS case~\cite{PolarizedTCS}, therefore we are replicating part of that setup.  Fig~\ref{fig:unpol_tcs_setup_full} shows our Geant4 simulations. The Compact Photon Source (CPS)~\cite{CPS1} will be used to generate a high intensity ($\sim10^{12} \gamma/sec$ 
circularly polarized photon beam. We are using a $15~cm$ long liquid hydrogen target, kept inside a scattering chamber. The outgoing particles are an electron, a positron and a recoil proton. To separate the outgoing $e^{-}$ and $e^{+}$, we placed  a magnet in front of the scattering chamber. Q1, Q2, Q3 and Q4 (in Fig.~\ref{fig:unpol_tcs_setup_full}) designate  
four identical "quadrants": each of the four quadrants consists of three layers of GEMs, hodoscopes and electromagnetic calorimeter. For further details about the base setup, see~\cite{PolarizedTCS}. The electromagnetic calorimeters in each quadrant correspond to $\sim1/2$ of the surface currently available from the Neutral Particle Spectrometer (NPS)~\cite{NPS}, currently 
used in a DVCS experiment  at JLab. 

\begin{figure}[!htb]
\centering
\includegraphics[width=0.7\textwidth,angle=0]{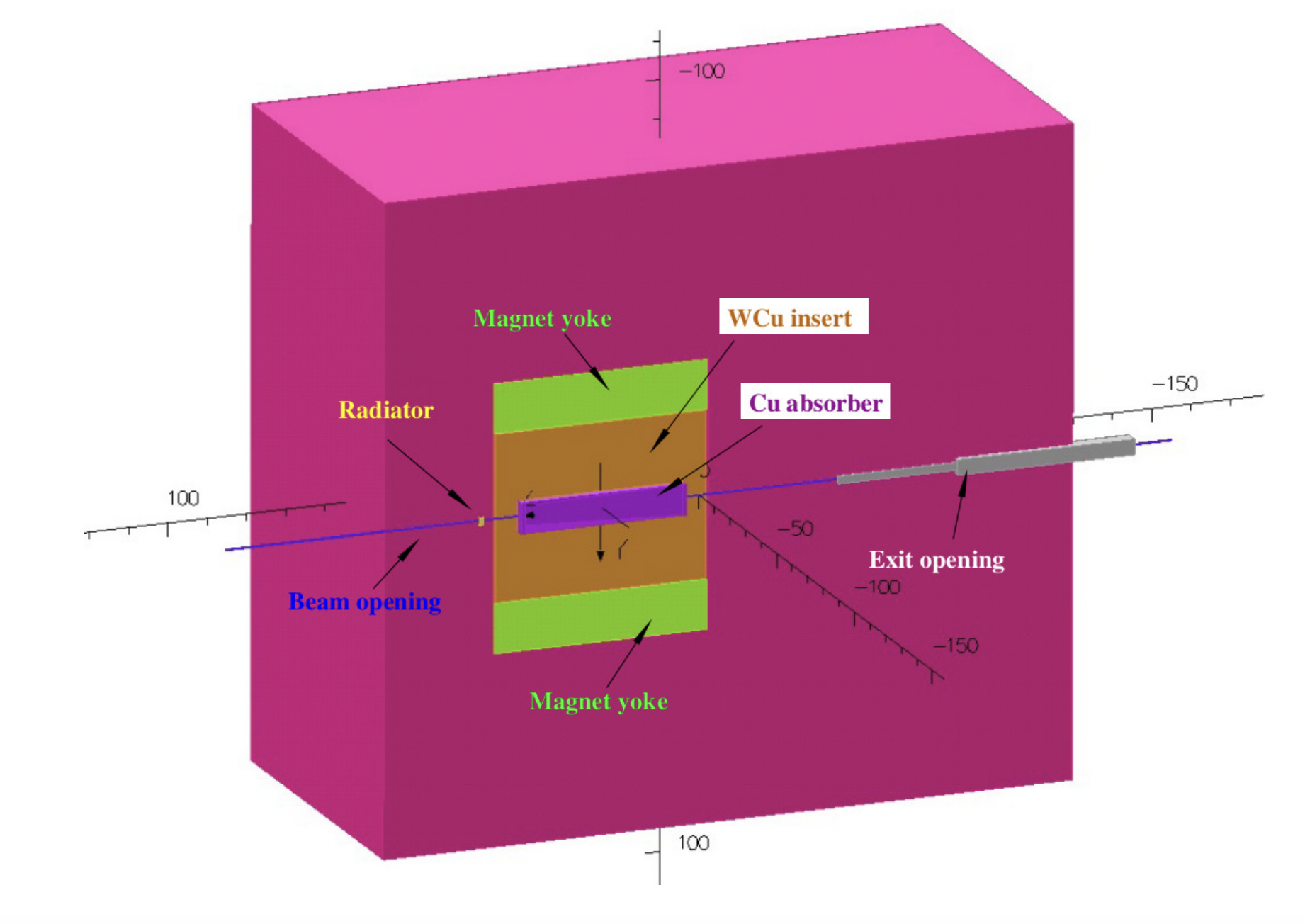}
\caption{The CPS cut-out side view. Deflected electrons strike a copper absorber, surrounded by a W-Cu
insert inside the magnet yoke. The outer rectangular region in this view is the tungsten-powder shield.}
\label{fig:compton_photon_source}
\end{figure}

\subsection{Compact Photon Source (CPS)}
The experiment will take advantage of the already approved Compact Photon Source (CPS) for the experiment E12-17-008~\cite{CPS,CPS1}, to get a pure circularly polarized  photon beam. The conceptual design of the CPS is shown in  Fig~\ref{fig:compton_photon_source}. The CPS can generate photons with a flux of $1.5\times 10^{12}~s^{-1}$ with a $2.5~\mu A$ electron beam, for photon energies $>$5~GeV. The expected spot size of the photon off the target is $\sim 1~mm$, at $2~m$ distance of the CPS radiator.  For more design details and working principle of CPS, we refer to~\cite{CPS_2}. 

\begin{figure}[!htb]
\centering
\includegraphics[width=0.7\textwidth,angle=0]{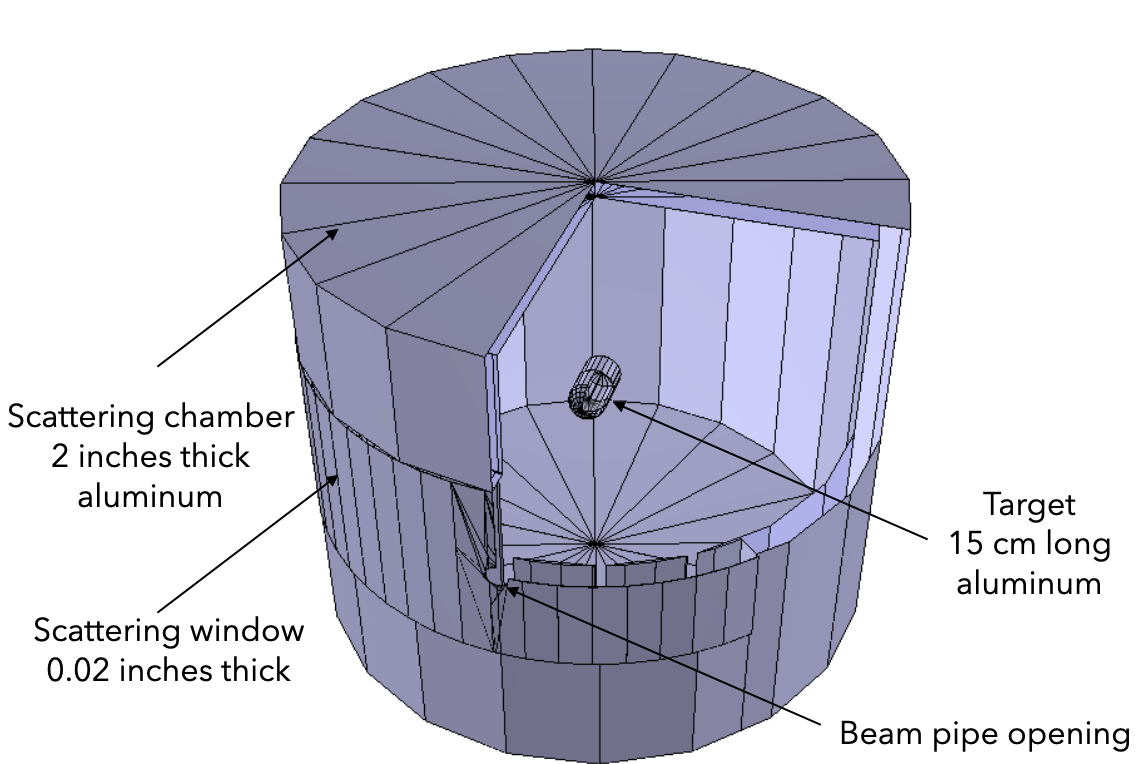}
\caption{The Geant4 design of the $15~cm$ long liquid hydrogen, target placed inside the scattering chamber.}
\label{fig:scattering_chamber_Geant4}
\end{figure}

\subsection{Target}
For this experiment, an unpolarized liquid hydrigen target is needed. A 15cm long liquid hydrogen target of Hall C is used in this study. The liquid hydrogen is kept in a $15~cm$ long aluminum can, and can be seen in the middle of the scattering chamber, in~Fig~\ref{fig:scattering_chamber_Geant4}. More details about the target geometry can be found in~\cite{StandardEquipmentManual}. The target is placed inside the scattering chamber, made of 2 inches thick aluminum wall. The horizontal angular range of the scattering window is from $3.2\deg$ to $77.0~\deg$ on the High Momentum Spectrometer (HMS, one of the "standard" Hall C detector equipment) side, and from $3.2\deg$ to $47.0~\deg$ on the Super High Momentum Spectrometer (SHMS, one of the "standard" Hall C detector equipment)  side. The thickness of the aluminum for the scattering window is $0.02~inches$.

\begin{figure}[!htb]
\centering
\includegraphics[width=0.5\textwidth,angle=0]{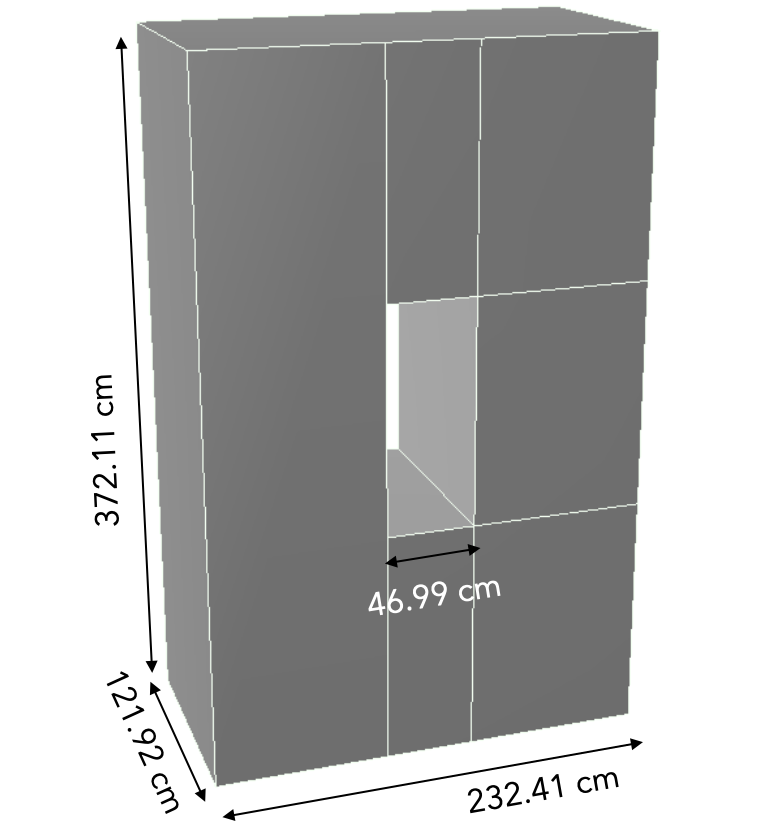}
\caption{GEANT4 simulation for SBS magnet geometry.}
\label{fig:SBS_magnet_Geant4}
\end{figure}

\subsection{Magnet}
A simple magnet geometry is simulated in Geant4 and placed between the scattering chamber and the detector stack~\ref{fig:SBS_magnet_Geant4}. 
The field strength is $2.4~T\text{-}m$ along the x axis (red arrow in Fig~\ref{fig:unpol_tcs_setup_full}), with $1.2~m$ long pole as shown in Fig~\ref{fig:SBS_magnet_Geant4}. 
What we eventually want will be to move the Super Big Bite Spectrometer (SBS) magnet from Jlab Hall A into  the Hall C, after their experiments. Therefore, we  is designed our magnet according to the SBS magnet geometry. We are looking into the possibility of reusing magnets from other experiments as well, to help in cutting the cost of our experiment.   

\subsection{GEMs}
For the reconstruction of the vertex parameters from the track coordinates at the detectors, tracker is needed for the experiment. For this purpose, we put 3 layers of GEMs in our simulations. The GEMs are very useful for the reconstruction of  transverse tracks with certainty up to $100~\mu m$. The general knowledge about  GEMs assures (compared to other methods of tracking) reliable track reconstruction in the background rates, which we expect to be relatively high (up to $10^{6} Hz/mm$ in this experiment). We found that  only 2 layers of the GEMs could be enough for the determination of  the positions and directions of the tracks, but we decided to put three layers to ensure it's performance in the high background rates and a better particle identification (see studies in~\cite{PolarizedTCS}).

\subsection{Hodoscopes}
The detection of recoil protons requires is essential in this experiment, because we are using an untagged photon beam from CPS, therefore are missing it's energy. Indeed, the scattered electron from the primary CEBAF beam is dumped into the CPS magnet. Our secondary photon beam ranges from $\sim5.5$ to $\sim11$ GeV. We know it's direction (very low scattering angle), but not it's exact energy. This is why, for an exclusive measurement such as TCS, we need to detect absolutely all the outgoing particles. The detection of the protons requires fly's-eye array of scintillators in each of the detector quadrants. The scintillators will be placed just before the NPS calorimeters, and the light sensors need to be coupled at the rear side of the scintillators. Our preliminary results show an optimal size for the scintillators at $2\times2\times5~cm^{3}$ for the transverse polarized case. We need to modify  this size for our unpolarized TCS experiment, since the magnetic field is different. 
The proton identifications will require the ${dE}/{dX}$ signal from the hodoscopes. The hodoscopes will also be used for the trigger system (from either the lepton pair solely, or all 3 scattered particles). For this experiment, the general idea of the trigger will follow the already proposed polarized TCS experiments. Details on the polarized case we are refering to are in~\cite{PolarizedTCS}.

\subsection{Electromagnetic calorimeters}
To determine the kinematic variables $Q^{\prime}$, $\xi$, and $t$, we need to measure the coordinates and the energy of the di-lepton ($e^{+}$, $e^{-}$) pair. We are using a highly segmented lead-tungstate calorimeter. As shown in the Fig~\ref{fig:unpol_tcs_setup_full}, each of the four detector quadrants has a calorimeter at the end. We plan to use the existing Neutral Particle Spectrometer (NPS) calorimeter (having enough crystals to prepare 2 of our "quadrants"). Our TCS experiment relies on doubling the surface of the NPS, i.e. the total number of crystals. In the  polarized TCS~\cite{PolarizedTCS} experimental setup, for each quadrant we have $23\times23$ matrix of PbWO4 crystal blocks. Each block has $2.05\times2.05~cm^{2}$ geometrical cross-sections.  

\begin{figure}[!htb]
\centering
\includegraphics[width=1.0\textwidth,angle=0]{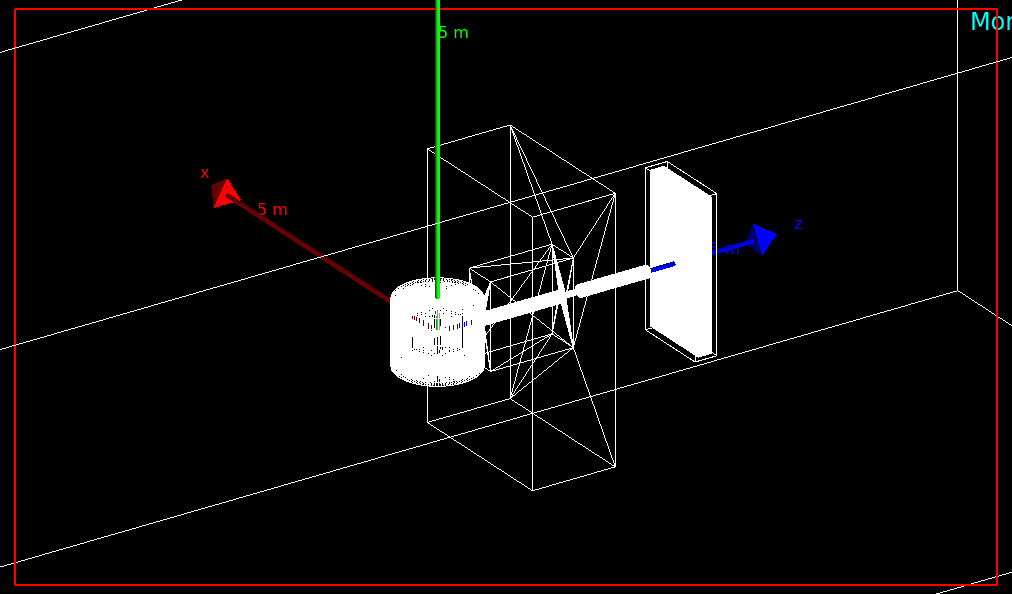}
\caption{Simple one calorimeter plane unpolarized TCS setup for the preliminary studies.}
\label{fig:TCS_unpol_setup}
\end{figure}

\section{Results}
At this stage of our studies,  we decided to study a simple one plane "calorimetric setup", and check the SBS magnet performances,  as shown Fig~\ref{fig:TCS_unpol_setup}. Here we removed all the detectors that were displayed Fig~\ref{fig:unpol_tcs_setup_full}, except for one single plane of calorimeter. The calorimeter plane is placed at the $90~\deg$ with $Z$ axis (parallel to $x\text{-}y$ plane), and facing the beam direction. The distance between the center of the target to the face of the calorimeter is $\sim 350~cm$. 

\begin{figure}[!htb]
     \centering
    \begin{subfigure}[b]{1.0\textwidth}
         \centering
         \includegraphics[width=\textwidth]{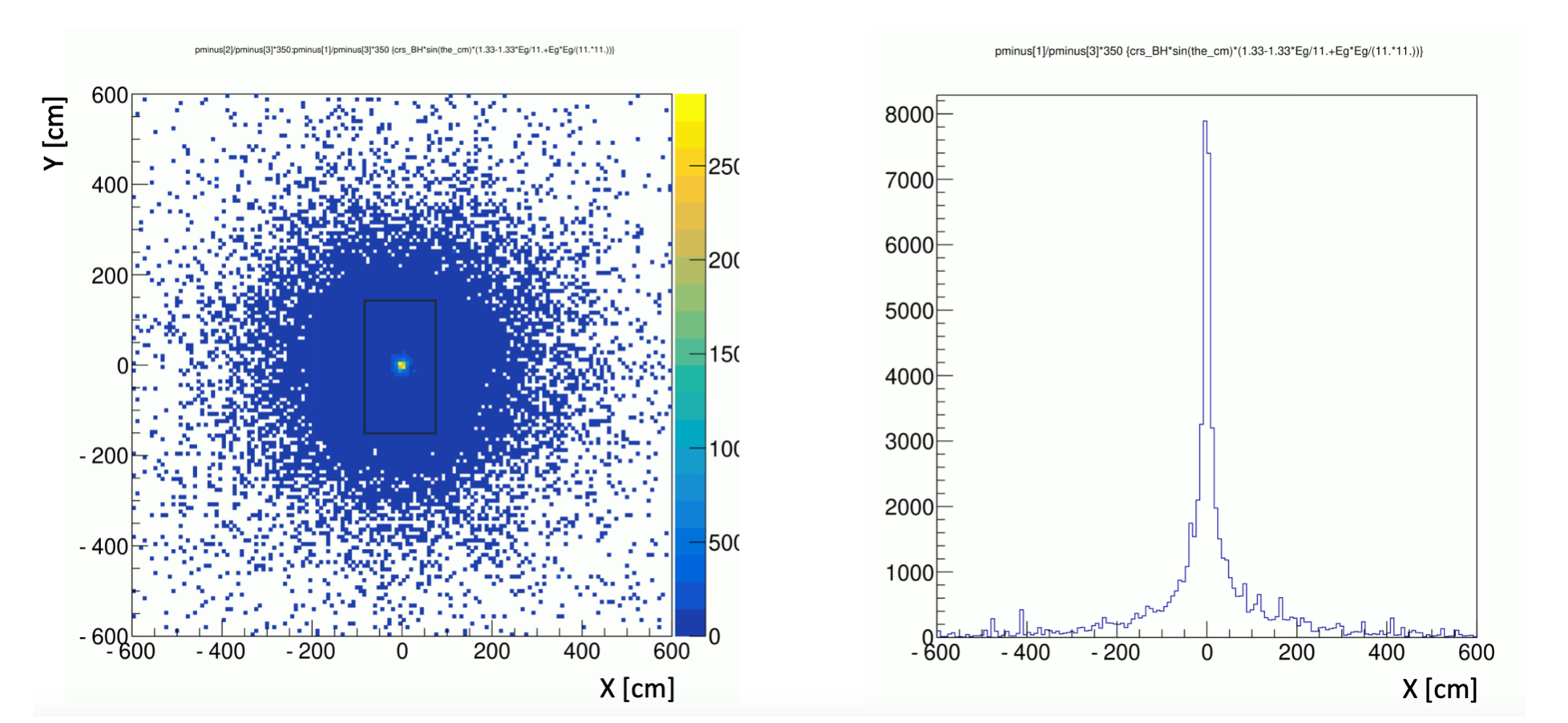}
         \caption{TCS weighted events for the for electrons, projected at the face of the calorimeter in Fig:\ref{fig:TCS_unpol_setup}. \textcolor{red}{No magnetic field} is used in the SBS magnet bore. The black box in the middle shows projected opening of the magnetic bore at the calorimeter face. Same is expected for the positrons. Figures are produced by Vardan Tadevosyan.}
         \label{fig:electron_projection_no_mag_field}
     \end{subfigure}
     \hfill
     \begin{subfigure}[b]{1.0\textwidth}
         \centering
         \includegraphics[width=\textwidth]{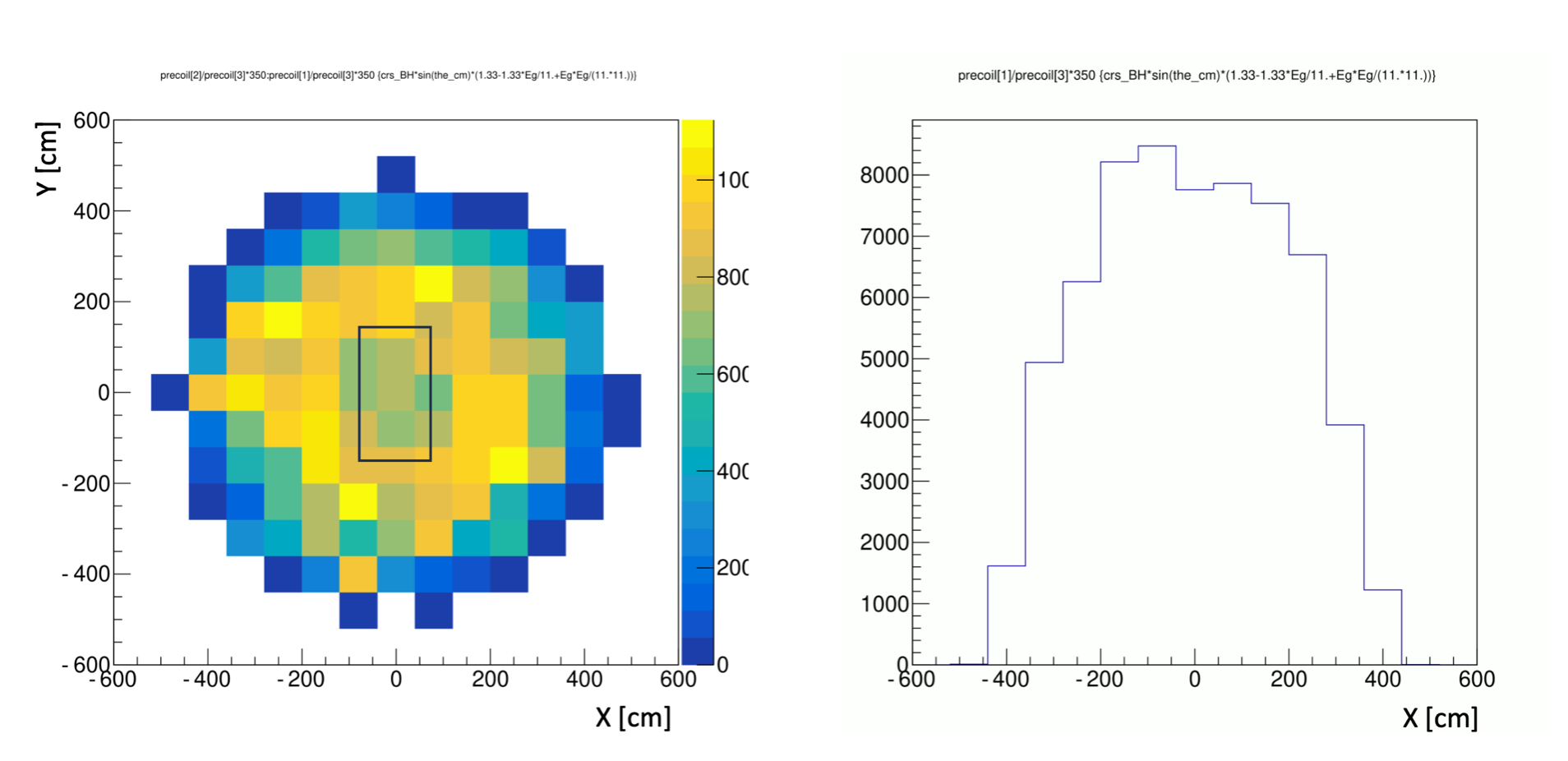}
         \caption{TCS weighted events for the for protons, projected at the face of the calorimeter in Fig:\ref{fig:TCS_unpol_setup}. \textcolor{red}{No magnetic field} is used in the SBS magnet bore. The black box in the middle shows projected opening of the magnetic bore at the calorimeter face. Figures are produced by Vardan Tadevosyan.}
         \label{fig:proton_projection_no_mag_field}
     \end{subfigure}
        \caption{Projected electron and the proton events at the face of the calorimeter (no magnetic field) in Fig~\ref{fig:TCS_unpol_setup}.}
        \label{fig:projection_no_mag_field}
\end{figure}

Without any magnetic field, the TCS weighted electron and proton events are projected at the face of the calorimeter. It is evident from Fig~\ref{fig:electron_projection_no_mag_field} that most of the electrons are well within the magnetic bore (black box), i.e. withing $\pm 50~cm$ in $x$ direction around the center of the calorimeter. So, the chance of SBS iron core blocking the electrons is quite low. As positrons are produced in pair production along with the electrons, we can expect the same positional distributions for the them at the face of the calorimeter. 

\begin{figure}[!htb]
     \centering
    \begin{subfigure}[b]{0.49\textwidth}
         \centering
         \includegraphics[width=\textwidth]{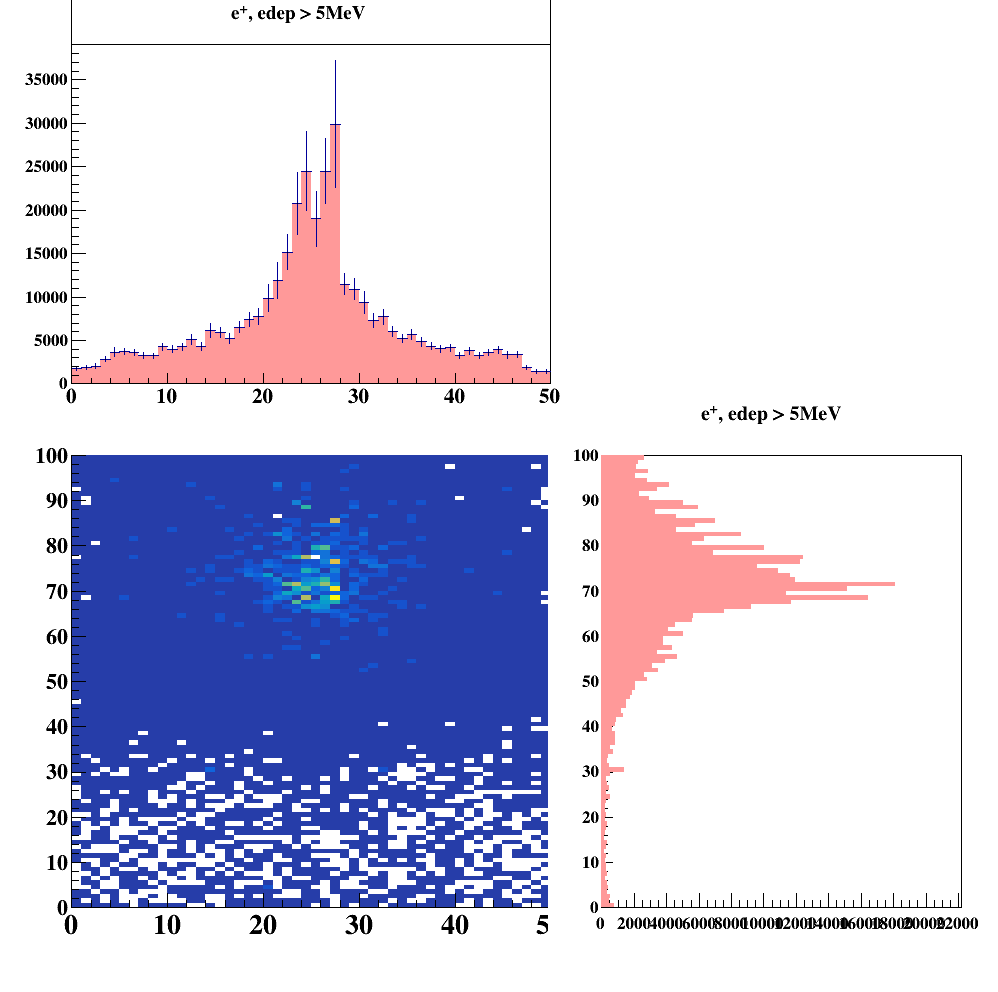}
         \caption{TCS weighted events for the for electrons, projected at the face of the calorimeter in Fig:\ref{fig:TCS_unpol_setup}. \textcolor{red}{$2.4 \ T\text{-}~m$ magnetic field} is used in the SBS magnet bore. The black box in the middle shows projected opening of the magnetic bore at the calorimeter face.}
         \label{fig:electron_projection_mag_field}
     \end{subfigure}
     \hfill
     \begin{subfigure}[b]{0.49\textwidth}
         \centering
         \includegraphics[width=\textwidth]{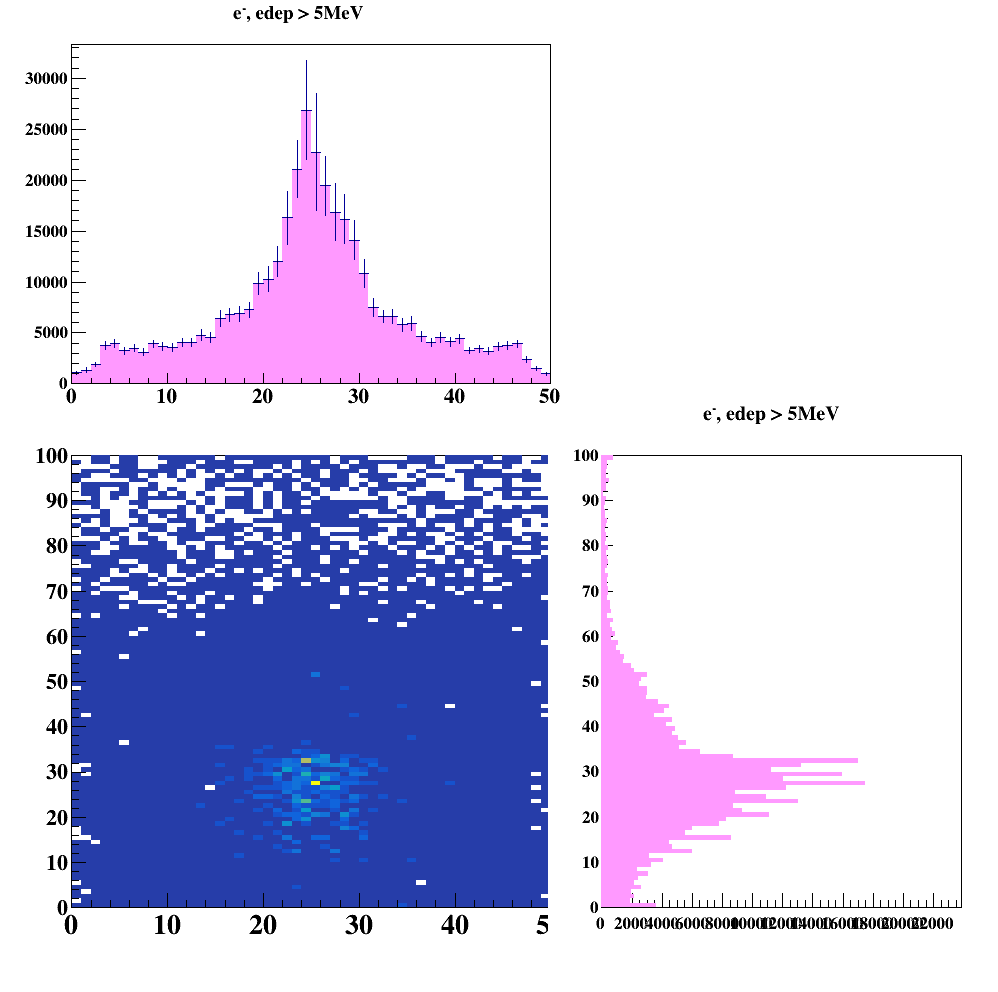}
         \caption{TCS weighted events for the for protons, projected at the face of the calorimeter in Fig:\ref{fig:TCS_unpol_setup}. \textcolor{red}{$2.4 \ T\text{-}~m$ magnetic field} is used in the SBS magnet bore. The black box in the middle shows projected opening of the magnetic bore at the calorimeter face. }
         \label{fig:positron_projection_mag_field}
     \end{subfigure}
        \caption{Projected electron and the proton events at the face of the calorimeter (with magnetic field) in Fig: \ref{fig:TCS_unpol_setup}.}
        \label{fig:projection_mag_field}
\end{figure}

After applying the magnetic field in the magnet, the electrons and positrons are separated and illuminated the different $y$ positions on the calorimeter as shown in the Fig~\ref{fig:electron_projection_mag_field} and Fig~\ref{fig:positron_projection_mag_field}. As expected there is no separation between the two particles in $x$ direction. Electrons populated the region between column 60 and column 80 of the scintillator segmentation, and the positrons populated the region between column 10 and 35. This clearly shows that our magnetic field simulation is acceptable, and with this setup the electrons and the positrons can be separated in space for particle identification.

As shown in the Fig~\ref{fig:proton_projection_no_mag_field}, even without the magnetic field the protons are not as concentrated in space as the electrons. In the $x$ direction the spread of the proton (without the magnetic field) is as big as $-400~cm$ to $+400~cm$ around the center of the calorimeter plane. Here we didn't show the spread of the protons with the magnetic field in the magnetic bore. 

\section{Future Work}
The work discussed in this paper is still progressing. In the next stage, we aim at getting realistic signal and background rates.
We will check if the current opening of the magnetic bore is big enough to detect a sufficient  number of protons  to conduct our experiment in reasonable amount of time. Furthermore, as we are following the basic detector detector geometry of the polarized TCS experiment, we are working together with the proponents of the "polarized case" to
 the detector design and calculate the full background (physics + experimental) contributions.   

\section{Summary}
We presented our setup
to measure unpolarized and  beam polarized (circularly) cross sections for the TCS+BH reactions  at Hall C in Jefferson Lab. 
Our work is still ongoing, but our simulatoions already show 
  that the SBS magnet can be used as a part of the PID process in the di-lepton spectrometer we are proposing. The basic design concept of the experimental setup is very similar to the polarized TCS experiment (already proposed), except the unpolarized target, circularly polarized beam and the SBS magnet added to the setup. We are currently working on optimizing the detectors and the full background simulations. Our goal is to submit a new proposal to the JLab Program Advisor Committee in 2024.

\end{document}